\documentclass[10pt,conference]{IEEEtran}
\usepackage{amsmath}
\usepackage{amssymb}
\usepackage{amsthm}
\usepackage{amsfonts}
\usepackage[dvips]{graphicx}
\usepackage{subfigure}

\usepackage{subfig}
\usepackage{algorithm}
\usepackage{algorithmic}

\usepackage{slashbox}
\usepackage{tabularx}
\usepackage{booktabs}
\usepackage{multirow}
\usepackage{cite}
\usepackage{color}
\usepackage{changepage}
\newtheorem{prop}{\textbf{Proposition}}[section]

\newtheorem{cor}{\textbf{Corollary}}

\newtheorem{lm}{\textbf{Lemma}}[section]

\newtheorem{thm}{\textbf{Theorem}}

\newcommand{\bthm}{\begin{thm}}
\newcommand{\ethm}{\end{thm}}

\newcommand{\bcor}{\begin{cor}}
\newcommand{\ecor}{\end{cor}}
\newcommand{\bprop}{\begin{prop}}
\newcommand{\eprop}{\end{prop}}
\newcommand{\blm}{\begin{lm}}
\newcommand{\elm}{\end{lm}}
\newcommand{\beq}{\begin{equation}}
\newcommand{\eeq}{\end{equation}}
\newcommand{\ber}{\begin{eqnarray}}
\newcommand{\eer}{\end{eqnarray}}

\newenvironment{proof1}{\begin{trivlist}\item[]{\bf Proof:\hspace{2mm}}}{\hfill$\blackbox$\end{trivlist}}

\newcommand{\argmax}{\mathop{\mbox{\rm arg\,max}}}

\newcommand{\blackbox}{\vrule height7pt width5pt depth1pt}

\newcommand{\bit}{\begin{itemize}}
\newcommand{\eit}{\end{itemize}}
\newcommand{\ben}{\begin{enumerate}}
\newcommand{\een}{\end{enumerate}}
\newcommand{\bdesc}{\begin{description}}
\newcommand{\edesc}{\end{description}}
\newcommand{\beqarrn}{\begin{eqnarray*}}
\newcommand{\eeqarrn}{\end{eqnarray*}}
\newenvironment{proofof}[1]{\begin{trivlist}\item[]{\bf Proof of #1:\hspace{2mm}
}}{\hfill\blackbox\end{trivlist}}
\newcommand{\bproofof}{\begin{proofof}}
\newcommand{\eproofof}{\end{proofof}}
\newenvironment{rem}{\begin{trivlist}\item[]{\bf
Remark:}\hspace{4mm}}{\end{trivlist}}
\newcommand{\brem}{\begin{rem}}
\newcommand{\erem}{\end{rem}}
\newenvironment{rems}{\begin{trivlist}\item[]{\bf
Remarks}\begin{itemize}}{\end{itemize}\end{trivlist}}
\newcommand{\brems}{\begin{rems}}
\newcommand{\erems}{\end{rems}}
\newtheorem{fact}{Fact}
\newcommand{\bfact}{\begin{fact}}
\newcommand{\efact}{\end{fact}}
\newtheorem{examp}{Example}
\newcommand{\bexamp}{\begin{examp}\rm}
\newcommand{\eexamp}{\end{examp}}
\newtheorem{defn}{Definition}
\newcommand{\bdefn}{\begin{defn}\rm}
\newcommand{\edefn}{\end{defn}}

\newtheorem{prob}{Problem}
\newcommand{\bprob}{\begin{prob}}
\newcommand{\eprob}{\end{prob}}

\newcommand{\bvtm}{\begin{verbatim}}
\newcommand{\bfig}{\begin{figure}}
\newcommand{\efig}{\end{figure}}
\newcommand{\bcen}{\begin{center}}
\newcommand{\ecen}{\end{center}}

\long\def\comment#1{}

\newcommand{\rev}[1]{{\color{black}#1}} 

\begin{document}
\title{Adaptive Network Coding for Scheduling Real-time Traffic with Hard Deadlines}

\author {\IEEEauthorblockN{Lei Yang\IEEEauthorrefmark{1},
Yalin Evren Sagduyu\IEEEauthorrefmark{2}, Jason Hongjun Li\IEEEauthorrefmark{2} and Junshan Zhang\IEEEauthorrefmark{1}}
\IEEEauthorblockA
{\IEEEauthorrefmark{1}School of ECEE,
Arizona State University, Tempe, AZ 85287, USA\\
}
\IEEEauthorblockA
{\IEEEauthorrefmark{2}Intelligent Automation, Inc., Rockville, MD 20855, USA \\
}
\IEEEauthorblockA
{Email: lyang55@asu.edu, ysagduyu@i-a-i.com, jli@i-a-i.com, junshan.zhang@asu.edu}
}
\maketitle
\begin{abstract}
We study adaptive network coding (NC) for scheduling real-time traffic over a single-hop wireless network.
To meet the hard deadlines of real-time traffic, it is  critical to strike a balance between maximizing the throughput and minimizing the risk that the entire block of coded packets may not be decodable by the deadline. Thus motivated, we explore  adaptive NC, where the block size is adapted based on the remaining time to the deadline, by casting this sequential block size adaptation problem as a finite-horizon Markov decision process. One interesting finding is that the optimal block size and its corresponding action space monotonically decrease as the deadline approaches, and the optimal block size is bounded by the ``greedy'' block size. These unique structures make it possible to narrow down the search space of dynamic programming, building on which we develop a monotonicity-based backward induction algorithm (MBIA) that can solve for the optimal block size in \emph{polynomial time}.
Since channel erasure probabilities would be time-varying in a mobile network,  we further develop a joint real-time scheduling and channel learning scheme with adaptive NC that can adapt to channel dynamics. We also generalize the analysis to multiple flows with hard deadlines and long-term delivery ratio constraints,  devise a low-complexity online scheduling algorithm integrated with the MBIA, and then establish its asymptotical throughput-optimality. In addition to analysis and simulation results, we perform high fidelity wireless emulation tests with real radio transmissions to demonstrate the feasibility of the MBIA in finding the optimal block size in real time.
\end{abstract}


\section{Introduction}
The past few years have witnessed a tremendous growth of multimedia applications in wireless systems. To support the rapidly growing demand in multimedia traffic, wireless systems must meet the stringent quality of service (QoS) requirements, including the minimum bandwidth and maximum delay constraints. However, the time-varying nature of wireless channels and the hard delay constraints give rise to great challenges in  scheduling  multimedia traffic flows.  In this paper, we explore \textit{network coding} (NC) to optimize the throughput of multimedia traffic over wireless \rev{channels} under the hard deadline constraint.

In capacitated multihop networks, NC is known to optimize the multicast flows from a single source to the min-cut capacity \cite{ahlswede:2000}. NC also provides coding diversity over unreliable wireless channels and improves the throughput and delay performance of single-hop broadcast systems, compared to (re)transmissions of uncoded packets \cite{ho:2006, eryilmaz:2008, medard:2008, nguyen:2009_2, yalin:2009, drinea:2009, barros:2009}.    Nevertheless, the block NC induces ``decoding delay,'' i.e., receivers may not decode network-coded packets until a sufficient number of innovative packets are received. Therefore, the minimization of NC delay has received much attention (e.g., \cite{traskov:2009,heide:2009,yeow:2009,yazdi:2009}).

For multimedia traffic, meeting the deadline may be more critical than reducing the average delay.
Under the \textit{hard deadline} constraints, NC may result in significant performance loss, unless the receivers can decode the packets before the deadline. Different NC mechanisms (e.g., \cite{wang:2010,nguyen:2009,eryilmaz:2010,eryilmaz:2011}) have been proposed recently to incorporate deadline constraints. An immediately-decodable network coding (IDNC) scheme has been proposed in \cite{wang:2010} to maximize the broadcast throughput subject to deadlines. A partially observable Markov decision process (POMDP) framework has been proposed in \cite{nguyen:2009} to optimize media transmissions with erroneous receiver feedback.

These works focus on optimizing network codes in each transmission; however, \rev{such an} approach is typically not tractable due to the ``curse of dimensionality'' of dynamic programming. To reduce the complexity of optimizing network codes in each transmission, \cite{eryilmaz:2010} has formulated a joint coding window selection and resource allocation problem to optimize the throughput in deadline-constrained flows. However, 
the computational complexity can be still overwhelming due to the finite-horizon dynamic programming involved in the coding window selection. To overcome this limitation, \cite{eryilmaz:2010} has proposed a heuristic scheme with fixed coding window to tradeoff between optimality and complexity.

A primary objective of this study is to (i) explore optimal adaptive NC schemes with low computational complexity, and (ii) integrate channel learning with adaptive NC over wireless broadcast erasure channels. Our main contributions are summarized as follows.
\begin{itemize}
\item We develop an adaptive NC scheme that sequentially adjusts the block size (coding block length) of NC to maximize the system throughput, subject to the hard deadlines (cf. \cite{eryilmaz:2010}). We show that the optimal block size and its corresponding action space monotonically decrease as the packet deadline approaches, and the optimal block size is bounded by the ``greedy'' block size \rev{that maximizes the immediate throughput only}. These unique structures make it possible to narrow down the search space of dynamic programming, and accordingly we develop a monotonicity-based backward induction algorithm (MBIA) that can solve for the optimal block size in \emph{polynomial time}, compared with \cite{nguyen:2009,eryilmaz:2010}.
    We also develop a joint real-time scheduling and \textit{channel learning} scheme with adaptive NC for the practical case, in which the scheduler does not have (perfect) channel information.

\item We generalize the study on adaptive NC  to the case with multiple flows. We develop a joint scheduling and block size adaptation approach to maximize the weighted system throughput subject to the long-term delivery ratio and the hard-deadline constraint of each flow. By integrating the MBIA in the model with multiple flows, we construct a low-complexity online scheduling  algorithm. This online algorithm is shown to be throughput optimal in the asymptotic sense as the step size in iterations approaches zero.

\item We implement the adaptive NC schemes in a realistic wireless emulation environment with real radio transmissions. Our high fidelity testbed results corroborate the feasibility of the MBIA in finding the optimal block size in real time. As expected, the adaptive NC scheme with the MBIA outperforms the fixed coding scheme, and the proposed scheme of joint real-time scheduling and channel learning performs well under unknown and dynamic channel conditions.
\end{itemize}

The rest of the paper is organized as follows. In Section \ref{sec:2}, we introduce the system model and present the block size adaptation problem with the hard deadlines. In Section \ref{sec:single frame}, we develop the MBIA to solve for the optimal block size and building on this we devise the joint real-time scheduling and channel learning scheme with adaptive NC for the case with unknown channel information. In Section \ref{sec:multiframe}, we generalize the study on adaptive NC to multiple flows. In Section \ref{sec:implementation}, we implement the adaptive NC schemes and test them in a realistic wireless emulation environment with hardware-in-the-loop experiments.
We conclude the paper in Section \ref{sec:conclusion}.

\section{Throughput Maximization vs. Hard Deadline}\label{sec:2}
\subsection{System Model}
\begin{figure}[tb!]
\begin{center}
\vspace{4cm}\hspace{2.2cm} {\includegraphics[scale=0.4]{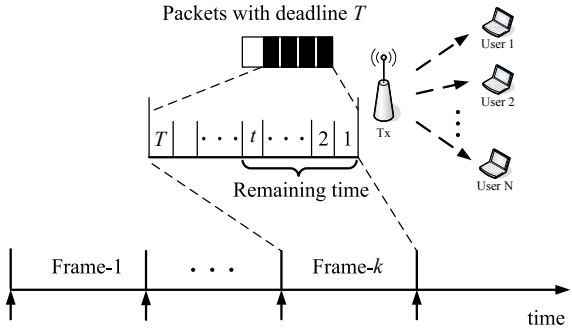}}\hspace{-0.0cm}
\vspace{0.6cm} \caption{System model. (The arrow denotes the time instant for drops of undelivered packets and arrivals of new packets.)}\vspace{-0.0cm}
\label{fig:model}
\end{center}
\end{figure}
We consider a time-slotted downlink system with one transmitter (e.g., base station) and $N$ receivers (users), as illustrated in Fig.~\ref{fig:model}. Time slots are synchronized across receivers and the transmission time of a packet corresponds to one time slot. The transmitter broadcasts $M$ packets to $N$ receivers over \textit{i.i.d.} binary erasure channels with erasure probability $\epsilon$.\footnote{The results derived in the paper can be readily applied to heterogeneous channels with different erasure probabilities.} We assume immediate and perfect feedback available at the transmitter. For multimedia communications, it is standard to impose deadlines for delay-sensitive data (see, e.g., \cite{nguyen:2009,wang:2010,drinea:2009,eryilmaz:2010,hou:2010}). We assume that packets must be delivered to each receiver before $T$ slots, i.e., the deadline of each packet is $T$ slots. Any packet that cannot be delivered to all receivers by this deadline is dropped without contributing to the throughput.

Worth noting is that this model can be readily applied to finite-energy systems with NC, where the objective is to maximize the system throughput before the energy is depleted for further transmission.
Therefore, the energy and delay constraints can be used interchangeably.

In Section \ref{sec:single frame}, we consider the basic model with one flow and one frame of $T$ slots. In Section \ref{sec:multiframe}, we generalize the model to multiple frames with multiple flows, where packets arrive at the beginning of each frame and they are dropped if they cannot be delivered to their receivers by the deadline of $T$ slots.

\subsection{Network Coding for Real-time   Scheduling}
As noted above, the throughput gain of NC comes at the expense of longer decoding delay (since packets are coded and decoded as a block), which may reduce the throughput of the system due to the hard deadline constraints. Let $K$ denote the block size, i.e., the number of original packets encoded together by NC. We assume that the transmitter and each receiver know the set of coding coefficients, and the transmitter broadcasts the value of $K$ to receivers before the NC transmissions start. The coding coefficients can also be chosen randomly from a large field size (or from a predetermined coding coefficient matrix of rank $K$) such that with high probability $K$ packet transmissions deliver $K$ innovative packets in coded form to any receiver, i.e., the entire block of packets can be decoded after $K$ successful transmissions. As shown in \cite{eryilmaz:2008}, the probability that all receivers can decode the block of size $K$ within $T$ slots is given by
\begin{equation}%
\begin{array}
[c]{lll}%
P(K,T)=\left(\sum\limits_{\tau=K}^T
 \binom{\tau-1}{K-1} \epsilon^{\tau-K}(1-\epsilon)^K\right)^N,
 \end{array}
\label{eq:P(K,T)}
\end{equation}
where $\binom{n}{m}$ denotes the number of combinations of size $m$ out of $n$ elements.\footnote{We can also employ random NC with a \rev{finite} field size $q$. This would change the decoding probability (\ref{eq:P(K,T)}) to a function of $q$. However, the general structure of the results will remain the same.} Note that (\ref{eq:P(K,T)}) strongly depends on the choice of block size $K$ and we can show that,

\begin{lm}\label{lm:P(K,T)}
The decoding probability (\ref{eq:P(K,T)}) is monotonically decreasing with $K$ \rev{for fixed $T$}.
\end{lm}

With block NC, there is the risk that none of the packets can be decoded by the receivers before the hard deadline. By using IDNC, it may be possible to start decoding without waiting for the entire block to arrive but the complexity of finding a suitable code may be overwhelming due to the dynamic programming involved in the problem \cite{wang:2010}. Here, we provide the throughput guarantees for the worst-case scenario, where either the whole block or none of the packets can be decoded at any slot. There is a tradeoff between the block size and the risk of decoding. In particular, we cannot greedily increase $K$ to maximize the system throughput under the hard deadline constraints, since the risk that some receivers cannot decode the packets, i.e., $1-P(K,T)$, also increases with $K$ according to Lemma \ref{lm:P(K,T)}.

If the first block is delivered within the deadline, i.e., $T$ slots, the size of a new block (with new packets) needs to be re-adjusted for the remaining slots. In other words, we need \textit{real-time} scheduling of network-coded transmissions depending on how close the deadline is. For example, when there is only one slot left before the deadline, the optimal block size is 1, since for any $K>1$, no receivers can decode the packets before the deadline. Also, the block size in a given slot statistically determines the remaining slots (before the deadline) along with the future system throughput. In Section \ref{sec:single frame}, we derive the optimal block size adaptation policy to maximize the system throughput under the deadline constraints for one frame with one flow. \rev{In Section} \ref{sec:multiframe}, we generalize the results to the case with multiple frames with multiple flows.

\subsection{Problem Formulation: A Markov Decision Process View}\label{sec:MDP}
The NC-based multimedia traffic scheduling of one frame is a sequential decision problem, which can be formulated as a Markov decision process (MDP) described as follows.

\emph{Horizon}: The number of slots available before the deadline over which the transmitter (scheduler) decides the block size is the horizon. Due to the hard deadline, this MDP problem is a finite horizon problem with $T$ slots (one frame).

\emph{State}: The remaining slots $t \in \{0,1,...,T \}$ before the hard deadline is defined as the state,\footnote{We use the terms ``state'' and ``slot'' interchangeably.} where $t=0$ denotes that there is no slot left for transmissions.

\emph{Action}: Let $K_t$, $ t \in \{1,...,T \}$, denote the action taken at state $t$, which is the block size for the remaining $t$ slots. Let $M_t$ denote the number of packets undelivered at state $t$. Thus, at state $t>0$, $K_t$ can be chosen from $1$ to $\min(t,M_t)$.
For $t=0$, the transmitter stops transmitting any packet, i.e., $K_0=0$. In general, the action space is defined as $\mathcal{K}_t = \{0,1,...,\min(t,M_t)\}$.

\emph{Expected immediate reward}: For the remaining $t$ slots, the expected immediate reward is the expected number of packets successfully decoded by all receivers, which is given by
\begin{equation}%
\label{eq:R_t}
\begin{array}
[c]{lll}%
R_t(K_t)=K_t \: P(K_t,t),
\end{array}
\end{equation}
where $P(K_t,t)$ is given by (\ref{eq:P(K,T)}), denoting the probability that each receiver can decode these $K_t$ packets within $t$ slots.

\emph{Block size adaptation policy}: A block size adaptation policy $\mathcal{P}$ is a sequence of mappings, $\mathcal{P}=\{\mathcal{P}_t\}_{t=1}^{T}$, from $t$, $M_t$, $\epsilon$, and $N$ to an action $K_t\in\{0,1,...,\min(t,M_t)\}$, i.e., $K_t=\mathcal{P}_t(t,M_t,\epsilon,N)=\min(\mathcal{P}_t(t,\epsilon,N),M_t)$. Without loss of generality, in Section \ref{sec:single frame}, we assume that $M_t$ is always larger than $t$, i.e., $K_t\in\{0,1,...,t\}$. This does not change the monotonicity structure of the block size with state $t$. We will discuss these structural properties in detail in Section \ref{sec:single frame}.


\emph{Total expected reward}: Given the adaptation policy $\mathcal{P}$, the total expected reward for the remaining $t$ slots is given by
\begin{equation}%
\begin{array}
[c]{lll}%
V_t(K_t;\mathcal{P})&=&R_t(K_t)+E[V_{j}(K_j;\mathcal{P})]\\
&=&R_t(K_t)+\sum\limits_{j=0}^{t-K_t}q_t(j)V_{j}(K_j;\mathcal{P}),
\end{array}
\end{equation}
where the probability mass function $q_t(j)=P(K_t,t-j)-P(K_t,t-j-1)$ denotes the probability that the block of size $K_t$ is delivered over exactly  $j$ slots before the deadline.

\section{Network Coding with Adaptive Block Size}\label{sec:single frame}
A main contribution of this paper is the development and analysis of the polynomial-time monotonicity-based backward induction algorithm (MBIA). The design of the MBIA is motivated by the structures of the optimal and the greedy policies that are formally defined as follows.

\begin{defn}
A real-time scheduling policy with adaptive network coding is optimal, if and only if it achieves the maximum value of the total expected reward given by the Bellman equation\cite{bertsekas:2005} in dynamic programming:
\begin{equation}%
\begin{array}
[c]{lll}%
V_t(K_t^*;\mathcal{P}^*)&=&\max\limits_{K_t\in\{0,1,...,t\}}\{R_t(K_t)\\&&+\sum\limits_{j=0}^{t-K_t}q_t(j)V_{j}(K_j^*;\mathcal{P}^*)\},
\end{array}
\label{eq:bellman}
\end{equation}
where $K_t^*$ denotes the optimal block size, $\mathcal{P}^*$ denotes the optimal block size adaptation policy, and the terminal reward is given by $V_0(0;\mathcal{P}^*)=0$.
\end{defn}

\begin{defn}
The greedy policy maximizes only the expected immediate reward (\ref{eq:R_t}) without considering the future rewards and the greedy decision is given by
\emph{
\begin{equation}%
\begin{array}
[c]{lll}%
\hat{K}_t=\argmax\limits_{K_t\in\{0,1,...,t\}} R_t(K_t).
\end{array}
\label{eq:greedy}
\end{equation}
}
\end{defn}

\subsection{Optimal Block Size Adaptation Policy}
In each slot $t$, the optimal policy balances the immediate reward and the future reward by selecting a suitable block size $K_t^*$. In general, the approach of solving for the optimal block size by traditional dynamic programming \cite{bertsekas:2005} suffers from the ``curse of dimensionality,'' where the complexity of computing the optimal strategy grows exponentially with $t$. However, the optimal block size and its corresponding action space exhibit the monotonicity structures, and the optimal block size is bounded by the greedy block size. These unique structures make it possible to narrow down the search space of dynamic programming, and accordingly we develop a monotonicity-based backward induction algorithm (MBIA) with polynomial time complexity.

The MBIA searches for the optimal block size by backward induction and provides the optimal block size for each system state. Depending on the remaining time to deadline, the scheduler transmits coded packets with the optimal block size until each user receives enough packets to decode this block. Then, the scheduler adjusts the block size based on the current state, and proceeds with the new block transmission. This continues until the packet deadline expires or all packets are delivered. We present next the structural properties of block size adaptation \rev{problem} that will lead to the formal definition of the MBIA.

\begin{lm}\label{lm:state space}
The action space $\mathcal{K}_t$ monotonically shrinks as $t$ decreases.
\end{lm}
\textit{Proof outline:} As the number of remaining slots $t$ decreases, the maximum possible block size decreases \rev{as well,} since $K_t\in\{0,1,...,t\}$; otherwise no \rev{receiver} can decode the block \rev{of coded packets}. \hfill$\square$

\begin{prop}\label{prop:unimodal}
The expected immediate reward function $R_t(K_t)$ has the following properties:
\begin{enumerate}
\item{$R_t(K_t)$ is unimodal for $K_t\in\{0,1,...,t\}$.\footnote{$f(x)$ is a unimodal function if for some $m$, $f(x)$ is monotonically increasing for $x \hspace{-0.075cm}\leq \hspace{-0.075cm} m$ and monotonically decreasing for $x \hspace{-0.075cm} \geq \hspace{-0.075cm} m$. The maximum value is attained at $x = m$ and there are no other local maximum points.}}
\item{$\hat{K}_t$ in (\ref{eq:greedy}) monotonically decreases as $t$ decreases.}
\end{enumerate}
\end{prop}
\textit{Proof outline:} To show the unimodal property, it suffices to show that $R_t(K_t)$ is log-concave, which can be shown by using induction method. The monotonicity property of $\hat{K}_t$ can be shown by invoking the contradiction argument and applying $\lim\nolimits_{t\rightarrow\infty}R_t(K_t)=K_t$. \hfill$\square$

\begin{figure}[t]
\begin{center}
\vspace{-0.0cm}\hspace{-0.0cm} {\includegraphics[scale=0.4]{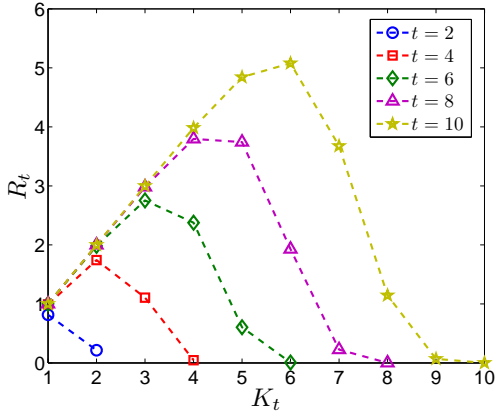}}\hspace{-0.0cm}
\vspace{-0.0cm} \caption{The unimodal property of $R_t(K_t)$.}\vspace{-0.0cm}
\label{fig:R_t}
\end{center}
\end{figure}

Fig.~\ref{fig:R_t} \rev{shows} the possible curves of $R_t(K_t)$ for different values of $t$, \rev{illustrating} the unimodal property of $R_t(K_t)$ \rev{formally stated in Proposition \ref{prop:unimodal}}.
Based on Proposition \ref{prop:unimodal}, the monotonicity property of the optimal block size $K_{t}^*$ is given by the following theorem.

\begin{thm}\label{thm:decreasing}
The optimal block size $K_t^*$ monotonically decreases as $t$ decreases, i.e., $K_{t}^*\ge K_{t-1}^*$, for any $t$.
\end{thm}
\textit{Proof outline:} Based on Proposition \ref{prop:unimodal}, we can show \rev{that} if $K_{t}^*<K_{t-1}^*$, $V_{t-1}(K_t^*;\mathcal{P}^*)>V_{t-1}(K_{t-1}^*;\mathcal{P}^*)$, which contradicts that $K_{t-1}^*$ is the optimal action in slot $t-1$. \hfill$\square$

As $t$ decreases, the risk that some receivers cannot decode the given block of packets increases for a fixed block size. Therefore, the scheduler becomes more conservative in the block size adaptation and selects a smaller block size.

{\bf{Remarks}:} 1) For $t=1,2$, the optimal block size is $K_t^*=1$, which can be obtained by computing the Bellman equation (\ref{eq:bellman}). 2) When $N=1$, the optimal block size is $K_t^*=1$ for all $t$, since the plain retransmission policy with $K_t = 1$ is better than the block coding with $K_t > 1$ in the presence of the hard deadlines.

\begin{thm}
The optimal block size $K_t^*$ is not greater than the greedy block size $\hat{K}_t$ for any $t$.
\label{thm:optimal<greedy}
\end{thm}
\textit{Proof outline:} Based on Proposition \ref{prop:unimodal}, we can show if $K_t^*>\hat{K}_t$, along any sample path, the system throughput by taking the action $\hat{K}_t$ is at least as high as that by taking the action $K_t^*$, which contradicts that $K_t^*$ in this case is the optimal action in slot $t$. \hfill$\square$

\begin{cor}\label{cor:Kt}
At state $t$, if $R_t(K_t)>R_t(K_t+1)$, then $K_j^*\le\hat{K}_j\le K_t$ for any $j\in\{1,...,t\}$.
\end{cor}

Corollary \ref{cor:Kt} follows \rev{directly} from Proposition \ref{prop:unimodal} and Theorem \ref{thm:optimal<greedy}.
Based on these structural properties, we develop the MBIA, which is presented in Algorithm \ref{alg:backward}.
\begin{algorithm}
\caption{Monotonicity-based Backward Induction Algorithm (MBIA)}
\label{alg:backward}
\begin{algorithmic}
\STATE 1) Set $t=0$ and $V_0(0;\mathcal{P}^*)=0$.
\STATE 2) Substitute $t+1$ for $t$, and compute $V_t(K_t^*;\mathcal{P}^*)$ by searching $K_t\in \mathcal{K}_t$, where $\mathcal{K}_t=\{K_{t-1}^*,K_{t-1}^*+1,...,\hat{K}_t\}$, i.e., $V_t(K_t^*;\mathcal{P}^*)=\max\limits_{K_t\in \mathcal{K}_t}\{R_t(K_t)+\sum\limits_{j=0}^{t-K_t}q(j)V_{j}(K_j^*;\mathcal{P}^*)\}$, and
$K_t^*=\argmax\limits_{K_t\in \mathcal{K}_t}\{R_t(K_t)+\sum\limits_{j=0}^{t-K_t}q(j)V_{j}(K_j^*;\mathcal{P}^*)\}$.
\STATE 3) If $t=T$, stop; otherwise go to step 2.
\end{algorithmic}
\end{algorithm}

\rev{The MBIA confines the search space at state $t$ to the interval from $K_{t-1}^*$ (the optimal policy at state $t-1$) to $\hat{K}_t$ (the greedy policy at state $t$).} \rev{Thus, the} MBIA reduces the search space over time and reduces the complexity of dynamic programming as given by the following theorem.
\begin{thm}\label{thm:complexity}
The MBIA is a polynomial-time algorithm and the complexity is upper bounded by $O(T^2)$.
\end{thm}
\textit{Proof:} Based on Proposition \ref{prop:unimodal}, $R_t(K_t)$ has the \emph{unimodal} property and therefore $\hat{K}_t$ can be solved efficiently by the Fibonacci search algorithm \cite{kiefer:1953}, which is a sequential line search algorithm with a complexity of $O(\log(t))$ at state $t$. Therefore, in each iteration, it takes $O(\log(t)+\hat{K}_t-K_{t-1}^*)$ slots to find $K_t^*$. Based on Lemma \ref{lm:state space}, $\hat{K}_t-K_{t-1}^*$ is upper bounded by $t$. After some algebra, we show that the complexity of Algorithm \ref{alg:backward} is bounded by $O(T^2)$ and Theorem \ref{thm:complexity} follows. \hfill$\square$

{\bf{Remarks:}} By using the MBIA, the optimal block size can be computed in polynomial time, which is a desirable property for online implementation. The optimal block size depends on the number of receivers and channel erasure probabilities. For different flows, the set of receivers may be different, which may result in different optimal block sizes, even when the number of remaining slots is the same across these flows. Therefore, without using the MBIA, offline schemes would need to compute the optimal policies for all possible receiver sets; however, this would be a computationally demanding task\rev{, as the number of receivers increases}.


Based on the monotonicity properties of the greedy and optimal block sizes, the optimal policy becomes the plain retransmission, if the channel erasure probability is sufficiently large. This sufficiency condition for $K^*_t=1$ at slot $t$ is formally given as follows.

\begin{thm}\label{thm:greedyoptimal}
At slot $t$, the optimal policy switches to the plain retransmission policy, i.e., $K^*_t = 1$, when the erasure probability satisfies the threshold condition
\begin{equation}%
\begin{array}
[c]{lll}%
\epsilon > \epsilon^*(t,N),
\end{array}
\label{eq:sufficient condition}
\end{equation}
where $\epsilon^*(t,N) \in (0,1)$ is the non-trivial (unique) solution to $R_t(1) = R_t(2)$.
\end{thm}
\textit{Proof outline:} The proof follows directly by comparing $R_t(1)$ and $R_t(2)$ that are expressed as a function of $\epsilon$. \hfill$\square$


Note that (\ref{eq:sufficient condition}) is a sufficient condition only and indicates the optimality of the greedy policy when $\epsilon$ is large enough.

Fig.~\ref{fig:threshold} depicts how the threshold $\epsilon^*$ varies with $t$ and $N$. The underlying monotonicity property is formally stated in Corollary \ref{cor:threshold}.
\begin{cor}
The threshold $\epsilon^*(t,N)$ increases monotonically with $t$ and decreases monotonically with $N$.
\label{cor:threshold}
\end{cor}

\begin{figure}[t]
\begin{center}
\vspace{-0.0cm}\hspace{-0.0cm} {\includegraphics[scale=0.4]{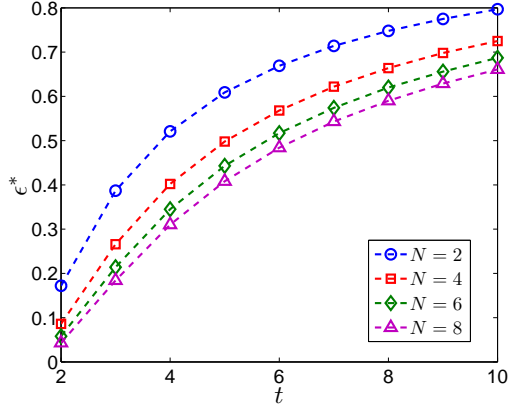}}\hspace{-0.0cm}
\vspace{0cm} \caption{The monotonicity property of $\epsilon^*$.}\vspace{-0.0cm}
\label{fig:threshold}
\end{center}
\end{figure}

{\bf{Remarks}:} 1) When the channel is good enough (with $\epsilon<\epsilon^*$), NC with $K_t > 1$ can always improve the throughput compared to the plain retransmission policy. 2) As $t$ increases (i.e., the deadline becomes looser), the risk of decoding network-coded packets decreases, i.e., $\epsilon^*(t,N)$ increases. 3) As $N$ increases, it becomes more difficult to meet the deadline for each of $N$ receivers and therefore $\epsilon^*(t,N)$ drops accordingly.

\subsection{Robustness vs. Throughput}
The real-time scheduling policies presented so far focus on the expected throughput without considering the variation from the average performance. Therefore, it is possible that the instantaneous throughput drops far below the expected value. To reduce this risk, we use additional variation constraints to guarantee that the throughput performance remains close to the average. In particular, for each slot $t$, we introduce the variation constraint \rev{to} the block size adaptation \rev{problem as follows}:
\begin{equation}%
\begin{array}
[c]{lll}%
v_t(K_t)<\sigma_t^2, ~\forall K_t\in \mathcal{K}_t,
\end{array}
\label{eq:var}
\end{equation}
where $\sigma_t^2$ is the maximum variation allowed in slot $t$ and the performance variation $v_t(K_t)$ under action $K_t$ is given by
\begin{equation}%
\begin{array}
[c]{lll}%
v_t(K_t)=\sum\limits_{i=1}^{\infty}i^2(P(K_t,i)-P(K_t,i-1)).
\end{array}
\label{eq:variation}
\end{equation}
Since $v_t(K_t)$ increases with $K_t$, (\ref{eq:var}) can be rewritten as the maximum block size constraint for each slot $t$, i.e.,
\begin{equation}%
\begin{array}
[c]{lll}%
K_t\le K_t^{\max},
\end{array}
\label{eq:Kmax}
\end{equation}
where $K_t^{\max}=\max\{K_t|K_t=\lfloor v_t^{-1}(\sigma_t)\rfloor\}$, $v_t^{-1}(\cdot)$ is the inverse mapping of $v_t(\cdot)$, and $\lfloor x\rfloor$ denotes the largest integer smaller than $x$. The variation constraints do not change the monotonicity property of the optimal block size provided by Theorem \ref{thm:decreasing}. By introducing the variation constraints (\ref{eq:Kmax}), the scheduler becomes more conservative in the block size adaptation.  The additional bound $K_t^{\max}$ can be easily incorporated into the MBIA by changing the action space to $\mathcal{K}_t=\{K_{t-1}^*,K_{t-1}^*+1,...,\min(K_t^{\max},\hat{K}_t)\}$ \rev{at state $t$}.

\subsection{Block Size Adaptation under Unknown Channels}
So far we have discussed the real-time scheduling policies with adaptive NC, where the channel erasure probability $\epsilon$ is perfectly known to the scheduler. The throughput performance of these policies depends on $\epsilon$; therefore, the scheduler needs to learn $\epsilon$ while adapting the block size, when it does not have (perfect) channel knowledge. Let $\hat\epsilon_t$ denote the estimate of the channel erasure probability in slot $t$. The scheduler can update $\hat\epsilon_t$ based on the feedback from the receivers. In slot $T$, if $\hat\epsilon_T<\epsilon$, we \rev{would} expect with high probability that a block of packets with the size that is calculated with respect to $\hat\epsilon_T$ cannot be delivered before the deadline. Therefore, it is better to select the block size conservatively at the beginning, when the estimate $\hat\epsilon_t$ cannot be highly accurate yet, because of the small number of samples. As $\hat\epsilon_t$ improves over time, the block size can be gradually increased to improve the system throughput. Once the estimate is close enough to the actual value of $\epsilon$ \rev{after enough samples are collected}, the block size should be adjusted (and reduced over time) according to the MBIA.

Clearly, there is a tradeoff between the channel learning and the block size adaptation. Here, we formulate a joint real-time scheduling and channel learning algorithm (Algorithm \ref{alg:learning}) to adapt the block size while updating the maximum likelihood estimate $\hat\epsilon_t$ of channel erasure probability. In slot $t$, based on the feedback, the scheduler can compute the packet loss ratio, \rev{$\epsilon_t=1-\frac{n_t}{N}$}, where \rev{$n_t$} denotes the number of \rev{receivers that successfully receive a packet in slot $t$}. Accordingly, the estimated channel erasure probability $\hat\epsilon_t$ is given by the moving average
\begin{equation}%
\begin{array}
[c]{lll}%
\hat\epsilon_t=\frac{(T-t)\hat\epsilon_{t+1}+\epsilon_t}{T-t+1}.
\end{array}
\label{eq:estimation}
\end{equation}
The scheduler decides on the block size by comparing the temporal variation $|\hat\epsilon_t-\hat\epsilon_{t+1}|$ with a threshold $\delta$. A detailed description is given in Algorithm \ref{alg:learning}.

\begin{algorithm}
\caption{Joint Real-time Scheduling and Channel Learning with Adaptive Network Coding}
\label{alg:learning}
\begin{algorithmic}
\STATE {\bf{Initialization:}} Choose threshold $\delta$ and set $K_T=1$.
\STATE {\bf{Repeat until}} $t=0$.
\STATE \quad  Update channel estimate $\hat\epsilon_t$ by (\ref{eq:estimation}).
\STATE \quad Compute block size $K_t^*$ by Algorithm \ref{alg:backward} with $\hat\epsilon_t$.
\STATE \quad \textbf{If} $|\hat\epsilon_t-\hat\epsilon_{t+1}|>\delta$ \textbf{then}
\STATE \qquad \textbf{If} $K_t^*\ge K_{t+1}+1$ \textbf{then}
\STATE \quad\qquad $K_t=K_{t+1}+1$,
\STATE \qquad \textbf{Else}
\STATE \quad\qquad $K_t=K_{t+1}$.
\STATE \qquad \textbf{Endif}
\STATE \quad \textbf{Else}
\STATE \qquad  $K_t=K_t^*$.
\STATE \quad \textbf{Endif}
\end{algorithmic}
\end{algorithm}

{\bf{Remarks:}} Algorithm \ref{alg:learning} captures the tradeoff between \rev{the} channel learning and block size adaptation. There are two options for the scheduler depending on \rev{the relationship between $|\hat\epsilon_t-\hat\epsilon_{t+1}|$ and} $\delta$. If the channel estimation is not yet good enough, Algorithm \ref{alg:learning} chooses the block size conservatively by incrementing $K_t$ by at most $1$. Otherwise, Algorithm \ref{alg:learning} computes the block size by applying the MBIA.

\subsection{Performance Evaluation}

Fig.~\ref{fig:K_t} illustrates for $N=5$ the monotonicity structure of the optimal block size (Theorem \ref{thm:decreasing}) and verifies that $K_t^*\le\hat{K}_t$ (Theorem \ref{thm:optimal<greedy}). Both the optimal and greedy block sizes increase when the channel conditions improve (from $\epsilon=0.5$ to $\epsilon=0.2$). Next, we evaluate the performance (average system throughput) of different policies. For comparison purposes, we also consider a soft delay-based \emph{conservative} policy, where the scheduler chooses the largest block size with the expected completion time less than or equal to the number of remaining slots. The expected completion time is studied in \cite{eryilmaz:2008}, and it is given by
\begin{equation}%
\begin{array}
[c]{lll}%
S(K)=K+\sum\limits_{t=K}^{\infty}\left(1-P(K,t)\right).
\end{array}
\label{eq:mean}
\end{equation}

Fig.~\ref{fig:comparison} compares the performance of the optimal, greedy, conservative and plain retransmission policies for $N=10$ and $T=10$. The plain retransmission policy always performs the worst, whereas the conservative policy performs worse than the greedy policy. \rev{However, as $\epsilon$ increases, all policies select smaller block sizes and their performance gap diminishes.}

Fig.~\ref{fig:variation} shows the tradeoff between the average system throughput and the throughput variation. When the channels are good (e.g., $\epsilon=0.1$ in Fig.~\ref{fig:variation}), the variation constraint (\ref{eq:var}) makes the scheduler choose a small block size, which reduces the average system throughput accordingly. However, there is no significant effect of (\ref{eq:var}) \rev{when channels are bad} (e.g., $\epsilon=0.5$ in Fig.~\ref{fig:variation}), since the scheduler already chooses a small block size for large $\epsilon$. Fig.~\ref{fig:learning} evaluates the performance of Algorithm \ref{alg:learning} under channel uncertainty and show that Algorithm \ref{alg:learning} is robust with respect to the variation of $\delta$ \rev{and achieves a reliable throughput performance close to the case with perfect channel information}.
\begin{figure}[t]
\begin{center}
\vspace{-0.0cm}\hspace{-0.0cm} {\includegraphics[scale=0.4]{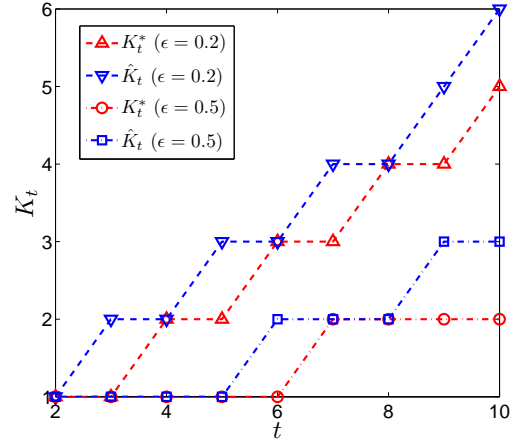}}\hspace{-0.0cm}
\vspace{-0.0cm} \caption{$K_t^*$ is nondecreasing and $K_t^* \leq \hat{K}_t$.}\vspace{-0.0cm}
\label{fig:K_t}
\end{center}
\end{figure}
\begin{figure}[t]
\begin{center}
\vspace{-0.0cm}\hspace{-0.0cm} {\includegraphics[scale=0.4]{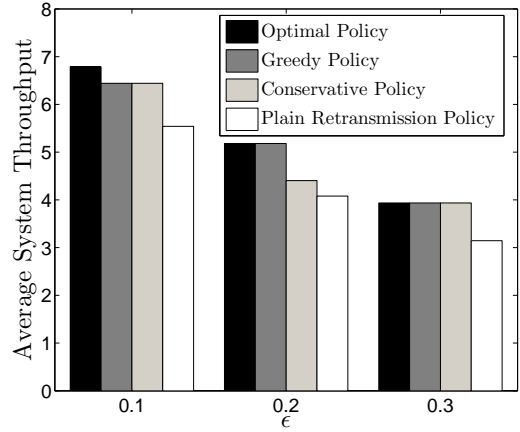}}\hspace{-0.0cm}
\vspace{-0.0cm} \caption{Performance (average system throughput) comparison of different policies.}\vspace{-0.0cm}
\label{fig:comparison}
\end{center}
\end{figure}
\section{Joint Scheduling and Block Size Optimization}\label{sec:multiframe}
In this section, we generalize the study on adaptive NC to the case of multiple frames, where the scheduler serves a set $\mathcal{F}$ of flows subject to the hard deadline and the long-term delivery ratio constraints. The packets of each flow $f$ arrive at the beginning of every frame and they are dropped if they cannot be delivered to its receivers $\mathcal{N}_f$ within this frame (see Fig.~\ref{fig:model}). We impose that the loss probability for flow $f$ due to deadline expiration must be no more than $1-q_f$, where $q_f$ is the delivery ratio requirement of flow $f$.
 For a given frame, the vector $a=(a_f)_{f\in\mathcal{F}}$  denotes the number of packet arrivals at each flow, where $a_f$ is the number of packets generated by flow $f$. We assume that $a_f$ is \emph{i.i.d.} across frames with finite mean $\lambda_f$ and variance\footnote{The algorithm developed for the \emph{i.i.d.} case can be readily applied to non \emph{i.i.d.} scenarios. \rev{The} analysis and performance guarantees can be obtained using the delayed Lyapunov drift techniques developed in \cite{neely:2006,neely:2010}.}. For ease of exposition, we \rev{assume perfect} channel information at the scheduler and consider coding within each flow but not across different flows.

\begin{figure}[t]
\begin{center}
\vspace{-0.0cm}\hspace{-0.5cm} {\includegraphics[scale=0.42]{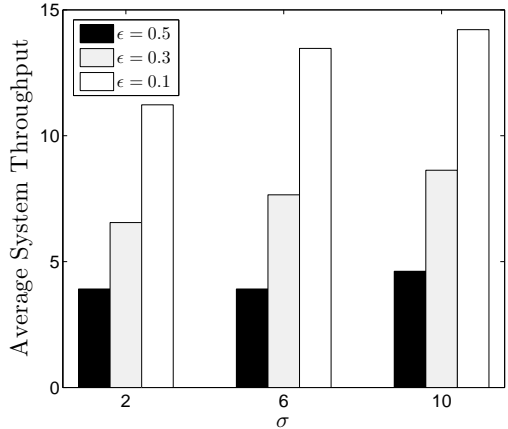}}\hspace{-0.0cm}
\vspace{-0.0cm} \caption{Average system throughput vs. throughput variation, where $N=10$ and $T=20$.}\vspace{-0.0cm}
\label{fig:variation}
\end{center}
\end{figure}

\begin{figure}[t]
\begin{center}
\vspace{-0.0cm}\hspace{-0.0cm} {\includegraphics[scale=0.42]{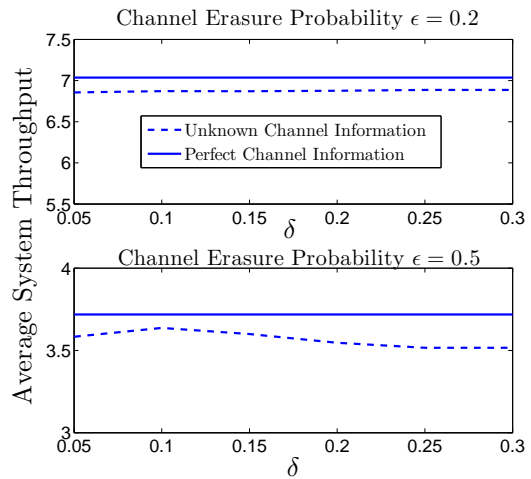}}\hspace{-0.0cm}
\vspace{-0.0cm} \caption{Performance of Algorithm \ref{alg:learning}.}\vspace{-0.0cm}
\label{fig:learning}
\end{center}
\end{figure}
\subsection{Multi-Flow Scheduling}
The scheduler allocates slots for each flow and uses the optimal real-time scheduling policy with adaptive NC developed in Section \ref{sec:single frame} to transmit network-coded packets. Given the arrivals, the scheduler needs to allocate a suitable number of slots for each flow to satisfy the delivery ratio requirement. This resource allocation is defined as a feasible schedule, $s=(s_f)_{f\in\mathcal{F}}$, where $s_f$ denotes the number of slots allocated to flow $f$ and $\sum_{f\in\mathcal{F}}s_f\le T$. Our goal is to maximize the weighted  throughput subject to the delivery ratio and hard deadline constraints. We find the optimal schedule, i.e., the probability $Pr(s|a)$ that given the arrivals $a$, the schedule $s\in\mathcal{S}$ is used from the set $\mathcal{S}$ of all feasible schedules. Then, the expected service rate for flow $f$ is upper-bounded by
\begin{equation}%
\begin{array}
[c]{lll}%
\mu_f\le \sum_{s, a} c_f(s_f) Pr(s|a)Pr(a),
\end{array}
\end{equation}
where $c_f(s_f)$ is the expected number of packets that can be delivered under schedule $s_f$, which is a constant and can be solved by the MBIA. Hence, we formulate the joint resource allocation and block size adaptation as the following optimization problem:
\begin{equation}%
\hspace{-0.2cm}
\begin{array}
[c]{lll}%
&{\text{maximize} }&\sum_{f\in\mathcal{F}}{w_f\mu_f}
\\
&\text{subject to}& \mu_f\ge \lambda_fq_f,  ~\forall~ f\in \mathcal{F},
\\
&& \mu_f\le \sum_{s, a} c_f(s_f) Pr(s|a)Pr(a),  ~\forall~ f\in\mathcal{F},
\\
&&Pr(s|a)\ge 0, ~\forall s\in\mathcal{S}, \:\: \sum_{s\in\mathcal{S}} Pr(s|a)\le 1, ~\forall a,
\\
& \text{variables}& \{\mu_f, Pr(s|a)\},
\end{array}
\label{eq:multiframe}
\end{equation}
where $w_f$ is the weight for flow $f$ and can be used as a fairness metric for resource allocation to each flow.\footnote{The problem (\ref{eq:multiframe}) can \rev{be generalized} to the case with congestion control by treating the weights as virtual queues for flow rates (similar to the service deficit queues that we use later in Section \ref{sec:dualdecomposition}).} Note that (\ref{eq:multiframe}) generalizes the problem studied in \cite{srikant:2011} by using adaptive NC schemes in packet transmissions.

\subsection{Dual Decomposition \label{sec:dualdecomposition}}
Since (\ref{eq:multiframe}) is strictly convex, the duality gap is zero from the Slater's condition \cite{boyd:2004}. The dual problem is given by
\begin{equation}%
\hspace{-0.2cm}
\begin{array}
[c]{lll}%
&{\text{maximize} }&\sum_{f\in\mathcal{F}}\left(w_f\mu_f+\nu_f(\mu_f-\lambda_fq_f)\right)
\\
&\text{subject to}& \mu_f\le \sum_{s, a} c_f(s_f) Pr(s|a)Pr(a),  ~\forall~ f\in\mathcal{F},
\\
&&Pr(s|a)\ge 0, ~\forall s\in\mathcal{S}, \:\: \sum_{s\in\mathcal{S}} Pr(s|a)\le 1,~\forall a,
\\
& \text{variables}& \{\mu_f, Pr(s|a)\},
\end{array}
\label{eq:multiframe_dual}
\end{equation}
where $\nu_f$ is the Lagrangian multiplier for flow $f$. The objective function of (\ref{eq:multiframe_dual}) is linear and the upper bounds for $\mu_f$ are affine functions. Therefore, the optimization problem (\ref{eq:multiframe_dual}) can be rewritten as:
\begin{equation}%
\begin{array}
[c]{lll}%
{\max\limits_{s\in\mathcal{S}} }\sum_{f\in\mathcal{F}}(w_f+\nu_f)c_f(s_f).
\end{array}
\label{eq:multiframe_dualdecoposition}
\end{equation}
Thus, we have the following gradient-based iterative algorithm to find the solution to the dual problem (\ref{eq:multiframe_dual}),
\begin{equation}%
\begin{array}
[c]{lll}%
s^*(k)\in{\argmax\limits_{s\in\mathcal{S}} }\sum_{f\in\mathcal{F}}(w_f+\nu_f(k))c_f(s_f),\\
\mu_f^*(k)=c_f(s_f^*(k)),\\
\nu_f(k+1)=\max(0,\nu_f(k)+\rho(\lambda_fq_f-\mu_f^*(k))),
\end{array}
\label{eq:iteration}
\end{equation}
where $k$ is the step index, $\rho>0$ is a fixed step-size parameter, and $c_f(s_f^*(k))$ is the expected service rate for flow $f$ under schedule $s_f^*(k)$.
Letting $\hat\nu_f(k)=\frac{\nu_f(k)}{\rho}$, (\ref{eq:iteration}) is rewritten as
\begin{equation}%
\begin{array}
[c]{lll}%
s^*(k)\in{\argmax\limits_{s\in\mathcal{S}} }\sum_{f\in\mathcal{F}}(\frac{w_f}{\rho}+\hat\nu_f(k))c_f(s_f),\\
\mu_f^*(k)=c_f(s_f^*(k)),\\
\hat\nu_f(k+1)=\max(0,\hat\nu_f(k)+(\lambda_fq_f-\mu_f^*(k))).
\end{array}
\label{eq:iteration1}
\end{equation}

{\bf{Remarks:}} The update equation for $\hat\nu_f$ can be interpreted as a virtual queue for the long-term delivery ratio with the arrival rate $\lambda_fq_f$ and the service rate $\mu_f^*(k)$, which keeps track of the deficit in service for flow $f$ to achieve a delivery ratio greater than or equal to $q_f$. Note that (\ref{eq:iteration1}) provides only the static solution to (\ref{eq:multiframe_dual}). Next, we provide an online scheduling algorithm which takes into account the dynamic arrivals of the flows.

\subsection{Online Scheduling Algorithm}
The online scheduling algorithm is given by
\begin{equation}%
\begin{array}
[c]{lll}%
\hspace{-0.25cm} s^*(k) \in {\argmax\limits_{s\in\mathcal{S}} }\sum_{f\in\mathcal{F}}(\frac{w_f}{\rho}+\hat\nu_f(k))c_f(s_f),\\
\hspace{-0.25cm} \hat\nu_f(k+1) = \max(0,\hat\nu_f(k)+\hat{a}_f(k)-\hat{c}_f(s_f^*(k))),
\end{array}
\label{eq:online}
\end{equation}
where $\hat{c}_f(s_f^*(k))$ denotes the actual delivered number of packets under the schedule $s_f^*(k)$ \rev{ depending} on the channel \rev{realizations}, and $\hat{a}_f(k)$ is a binomial random variable with parameters $a_f(k)$,  the number of packet arrivals of flow $f$ in the $k$th frame, and $q_f$. This implementation for $\hat{a}_f(k)$ was proposed in \cite{srikant:2011}. At the beginning of each \rev{period,} the schedule $s^*(k)$ is determined by (\ref{eq:online}). Then, the packets of each flow $f$ are transmitted with the MBIA in the scheduled $s^*_f(k)$ slots. The virtual
queue $\hat\nu_f$ is updated based on the number of successfully delivered packets $\hat{c}_f(s_f^*(k))$ of each flow $f$.
With Lyapunov optimization techniques \cite{neely:2006,neely:2010}, it can be shown that (\ref{eq:online}) has the following properties.

\begin{thm}
Consider the Lyapunov function $L(\hat\nu)$ $=\frac{1}{2}\sum_{f\in\mathcal{F}}\hat\nu_f^2$. If $\mu_f^*> \lambda_fq_f$ for all $f\in\mathcal{F}$, then the expected service deficit $\hat\nu_f$ is upper-bounded by
\begin{equation}%
\begin{array}
[c]{lll}%
\limsup\limits_{k\rightarrow\infty}E[\sum_{f\in\mathcal{F}}\hat\nu_f(k)]\le B_1+\frac{1}{\rho}B_2,
\end{array}
\nonumber
\end{equation}
for some positive constants $B_1$ and $B_2$. Furthermore, the online algorithm can achieve the long-term delivery ratio requirements, i.e., for all $f\in\mathcal{F}$ we have
\begin{equation}%
\begin{array}
[c]{lll}%
\liminf\limits_{K\rightarrow\infty}E[\frac{1}{K}\sum_{k=1}^{K}\hat{c}_f(s_f^*(k))]\ge \lambda_fq_f.
\end{array}
\nonumber
\end{equation}
\end{thm}

\begin{thm}
Let $\rho>0$ and $\mu_f^*$ be the solution to (\ref{eq:iteration1}). If $\mu_f^*> \lambda_fq_f$ for all $f\in\mathcal{F}$, it follows for $B>0$ that
\begin{equation}%
\begin{array}
[c]{lll}%
\limsup\limits_{K\rightarrow\infty}E[\sum_{f\in\mathcal{F}}(w_f\mu_f^*
-\frac{w_f}{K}\sum_{k=1}^{K}\hat{c}_f(s_f^*(k)))]\le B\rho.
\end{array}
\nonumber
\end{equation}
\end{thm}

The proofs follow from the optimization framework in \cite{neely:2006,neely:2010} and they are similar to the proofs presented in \cite{srikant:2011}.
Note that  the online scheduling algorithm (\ref{eq:online}) can approach within $O(\rho)$ of the optimal solution to (\ref{eq:multiframe_dual}) and does not require any knowledge of \rev{the packet arrival statistics.}

\subsection{Performance Evaluation}
We consider a network with two flows, each with five receivers. The packet traffic of each flow follows Bernoulli distribution with mean $\lambda_f$ packets/frame for $f=1,2$, and the length of each frame is $10$ slots. In the simulation, we set $\lambda_f=\lambda$ for $f=1,2$. The channel erasure probability $\epsilon$ is $0.3$, the weights $w_f$ are $1$ for all flows, the step-size $\rho$ is $0.1$, and the simulation time is $10^5$ frames.

We evaluate the performance of our algorithm by comparing the region of achievable rates $(\mu_1,\mu_2)$ with the plain retransmission under different traffic flow rates $\lambda$, where the achievable rates denote the feasible solution to (\ref{eq:multiframe}) for given delivery ratio requirements $q_f$. By \rev{varying $q_f$}, we find the achievable rate region. As illustrated in Fig.~\ref{fig:stability}, the plain retransmission only achieves a small fraction of the region with adaptive NC. By using adaptive NC, the network can support \rev{flows with heavier traffic}.

Fig.~\ref{fig:multiframe} shows the average service deficit $\hat\nu$ of two flows. The delivery ratio requirement of each flow is $0.8$. As $\lambda$ increases, $\hat\nu$ grows unbounded, which means that the conditions, $\mu_f^*> \lambda_fq_f$ for all $f\in\mathcal{F}$, are not satisfied, i.e., the arrival rates are not in the ``stability'' region, and the online scheduling algorithm cannot meet the delivery ratio requirements.

\begin{figure}[tb!]
\centering
\subfigure[$\lambda=3$ (packets/frame)]{\includegraphics[scale=0.42]{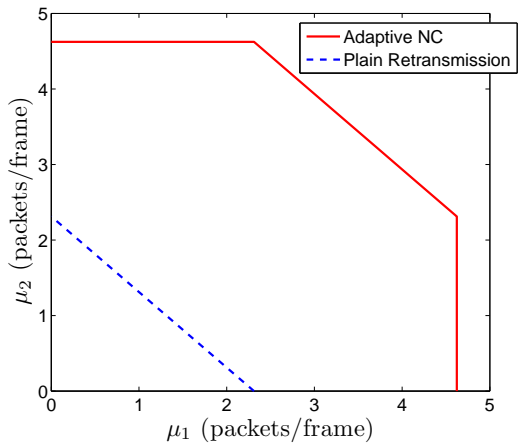}}
\subfigure[$\lambda=4.8$ (packets/frame)]{\includegraphics[scale=0.42]{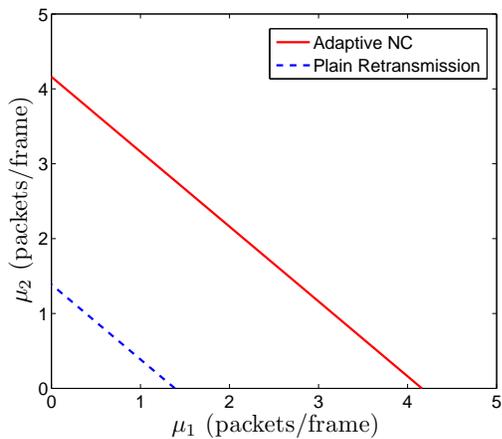}}
\caption{Achievable rate regions under adaptive NC and plain retransmission policies.}\label{fig:stability}
\end{figure}

%

\begin{figure}[t]
\begin{center}
\vspace{-0.0cm}\hspace{-0.0cm} {\includegraphics[scale=0.4]{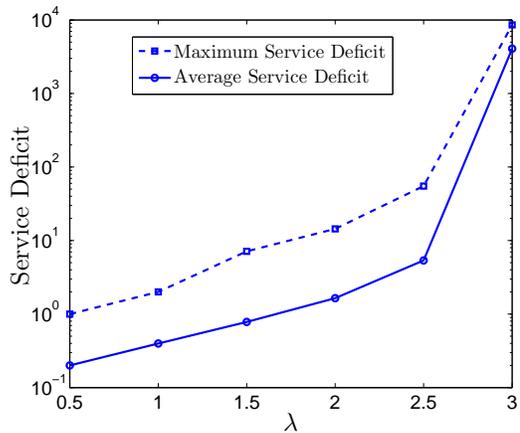}}\hspace{-0.0cm}
\vspace{-0.0cm} \caption{Service deficit vs. average arrival rate.}\vspace{-0.0cm}
\label{fig:multiframe}
\end{center}
\end{figure}

\section{High Fidelity Wireless Testing with Hardware Implementation} \label{sec:implementation}
We tested the adaptive NC schemes in a realistic wireless emulation environment with real radio transmissions. \rev{As illustrated in Fig.~\ref{fig:RFnest}, our} testbed platform consists of four main components: radio frequency network emulator simulator tool, RFnest\texttrademark\cite{iai:2011} (developed and owned as a trademark by Intelligent Automation, Inc.), software simulator running higher-layer protocols on a PC host, configurable RF front-ends (RouterStation Pro from Ubiquiti), and digital switch. We removed the radio antennas and connected the radios with RF cables over an attenuator box. Then, real signals are sent over emulated channels, where actual physical-layer interactions occur between radios, and in the meantime the physical channel attenuation is digitally controlled according to \rev{the simulation model or recorded field test scenarios can be replayed.}

In the hardware experiments, we executed wireless tests at 2.462GHz channel with 10dBm transmission power and 1Mbps rate. We used CORE (Common Open Research Emulator) \cite{core:2008} to manage the scenario being tested. We changed the locations of receivers through RFnest\texttrademark GUI and \rev{let the signal power decay as $d^{-\alpha}$ over distance $d$ with path loss coefficient $\alpha=4$. By using real radio transmissions according to this model, we varied the attenuation from the transmitter to each of the receivers and generated different channel erasure probabilities. With RFnest\texttrademark,} we replayed the same wireless traces for each of the NC algorithms and compared them under the high fidelity network emulation with hardware-in-the-loop experiments.

Fig.~\ref{fig:comparsion implementation} illustrates the performance of the optimal policy, the greedy policy and the fixed block size policy suggested by \cite{eryilmaz:2010}. The experimental results show that the greedy policy performs close to the optimal policy in practice. Both the greedy and the optimal policies outperform the fixed block size policy, and the complexity remains low with the polynomial-time algorithm MBIA. Fig.~\ref{fig:unknow} illustrates the wireless test performance for the case when the unknown channel erasure probabilities are learned over time. Algorithm \ref{alg:learning} performs close to optimal in this case and converges quickly in several frames.

\begin{figure}[t]
\begin{center}
\vspace{0.5cm}
\hspace{-0.2cm}
\includegraphics[scale=0.7]{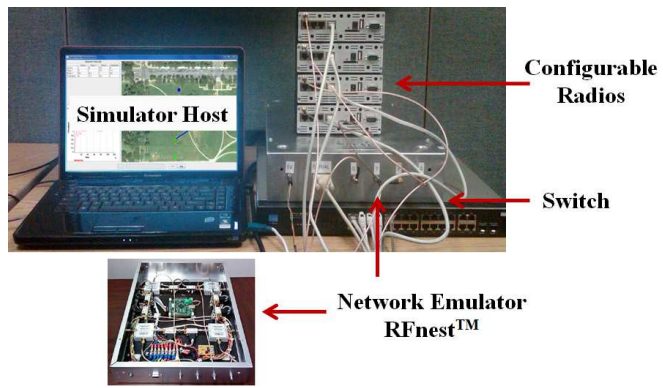}
\vspace{-0.0cm} \caption{Programmable RFnest\texttrademark testbed.}\vspace{-0.5cm}
\label{fig:RFnest}
\end{center}
\end{figure}

\begin{figure}[t]
\begin{center}
\vspace{-0.0cm}\hspace{-0.8cm} {\includegraphics[scale=0.42]{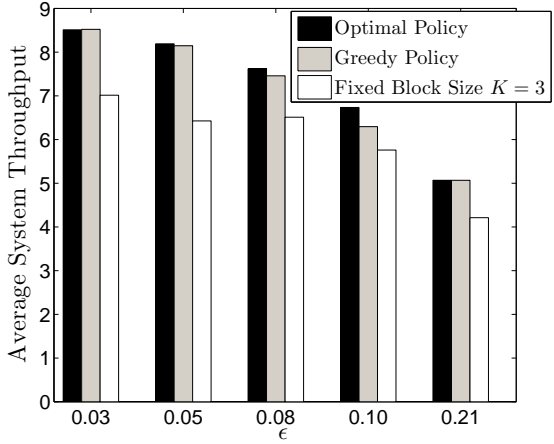}}\hspace{-0.0cm}
\vspace{-0.0cm} \caption{Performance comparison of different NC block size adaptation policies with network emulation, where $N=10$ and $T=10$.}\vspace{-0.0cm}
\label{fig:comparsion implementation}
\end{center}
\end{figure}

\begin{figure}[t]
\begin{center}
\vspace{-0.0cm}\hspace{-0.0cm} {\includegraphics[scale=0.45]{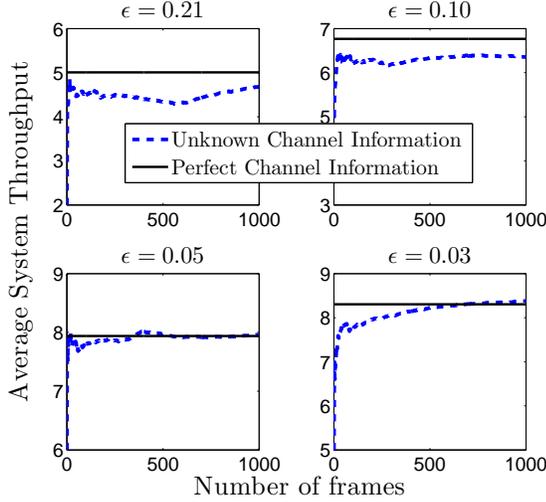}}\hspace{0.0cm}
\vspace{-0.3cm} \caption{Average system throughput and convergence rate of Algorithm \ref{alg:learning}, where $N=10$, $T=10$ and the initial channel erasure probability estimation is 0.5.}\vspace{-0.0cm}
\label{fig:unknow}
\end{center}
\end{figure}

\section{Conclusion}\label{sec:conclusion}
We considered adaptive NC for multimedia traffic with hard deadlines and formulated the sequential block size adaptation problem as a Markov decision process for a single-hop wireless network. By exploring the structural properties of the problem, we derived the polynomial time policy, MBIA, \rev{to solve} the optimal NC block size adaptation \rev{problem} and developed the joint real-time scheduling and channel learning scheme that can adapt to wireless channel dynamics if the perfect channel information is not available at the scheduler. Then, we generalized the study to multiple flows with hard deadlines and long-term delivery constraints, and developed a low-complexity online scheduling algorithm integrated with the MBIA. Finally, we performed high fidelity wireless emulation tests with real radios to demonstrate the feasibility of the MBIA in finding the optimal block size in real time.
Future work should extend the model \rev{to integrate congestion control with adaptive NC and real-time scheduling under deadline constraints.}
\section*{Acknowledgments}
We would like to thank Lei Ding from Intelligent Automation, Inc. for help with hardware experiments. This material is based upon work supported by the Air Force Office of Scientific Research under Contracts FA9550-10-C-0026, FA9550-11-C-0006 and FA9550-12-C-0037. Any opinions, findings and conclusions or recommendations expressed in this material are those of the authors and do not necessarily reflect the views of the Air Force Office of Scientific Research .

\appendix
\subsection{Proof of Lemma \ref{lm:P(K,T)}}\label{proof:P(K,T)}
$P(K,T)$ is monotonically decreasing with $K$, if $\hat{P}(K,T)$ is monotonically decreases with $K$, where
\begin{equation}%
\begin{array}
[c]{lll}%
 \hat{P}(K,T)=\sum\limits_{\tau=K}^T
 \binom{\tau-1}{K-1} \epsilon^{\tau-K}(1-\epsilon)^K.
\end{array}
\label{eq:P(K,T) N=1}
\end{equation}
such that $P(K,T) = (\hat{P}(K,T))^N$. First, we express
\begin{equation}%
\begin{array}
[c]{lll}%
&&\hat{P}(K,T)-\epsilon\hat{P}(K,T)\\
&=&(1-\epsilon)^K+\sum\limits_{\tau=K+1}^T\left(\binom{\tau-1}{K-1}-\binom{\tau-2}{K-1}\right)\epsilon^{\tau-K}(1-\epsilon)^K\\
&&-\binom{T-1}{K-1}\epsilon^{T-(K-1)}(1-\epsilon)^{K}\\
&=&(1-\epsilon)^K+\sum\limits_{\tau=K+1}^T\binom{\tau-2}{K-2}\epsilon^{\tau-K}(1-\epsilon)^K\\
&&-\binom{T-1}{K-1}\epsilon^{T-(K-1)}(1-\epsilon)^{K}\\
&=&(1-\epsilon)\left(\sum\limits_{\tau=K}^T\binom{\tau-2}{K-2}\epsilon^{\tau-K}(1-\epsilon)^{K-1}\right)\\
&&-\binom{T-1}{K-1}\epsilon^{T-(K-1)}(1-\epsilon)^{K}\\
&=&(1-\epsilon)\Big(\sum\limits_{\tau=K-1}^T\binom{\tau-1}{K-2}\epsilon^{\tau+1-K}(1-\epsilon)^{K-1}\\
&&-\binom{T-1}{K-2}\epsilon^{T+1-K}(1-\epsilon)^{K-1}\Big)\\
&&-\binom{T-1}{K-1}\epsilon^{T-(K-1)}(1-\epsilon)^{K}\\
&=&(1-\epsilon)\hat{P}(K-1,T)-\binom{T-1}{K-2}\epsilon^{T+1-K}(1-\epsilon)^{K}\\
&&-\binom{T-1}{K-1}\epsilon^{T-(K-1)}(1-\epsilon)^{K}\\
&&=(1-\epsilon)\hat{P}(K-1,T)-\binom{T}{K-1}\epsilon^{T-(K-1)}(1-\epsilon)^{K}.
\end{array}
\label{eq:P(K,T)expansion}
\end{equation}

Since from (\ref{eq:P(K,T)expansion}) it follows that
\begin{equation}%
\begin{array}
[c]{lll}%
\hspace{-0.4cm} \hat{P}(K,T) \hspace{-0.1cm} - \hspace{-0.1cm} \hat{P}(K\hspace{-0.1cm}-\hspace{-0.1cm}1,T)\hspace{-0.1cm}= \hspace{-0.1cm}-\binom{T}{K-1}\epsilon^{T-(K-1)}(1-\epsilon)^{K-1}
\end{array}
\label{eq:P(K,T) difference}
\end{equation}
is negative, $P(K,T)$ is monotonically decreasing with $K$.

\subsection{Proof of Proposition \ref{prop:unimodal}}\label{proof:unimodal}
1) To show that $R_t(K_t)$ is unimodal, it suffices to show that $R_t(K_t)$ is log-concave, i.e., $\hat{R}(K)=\frac{1}{N}\log(K)+\log(\hat{P}(K,T))$ is concave. Since $\frac{1}{N}\log(K)$ is concave, it suffices to show that for any given $T$, $\log(\hat{P}(K,T))$ is concave, i.e., $\hat{P}(K,T)$ is log-concave. Based on the definition of log-concavity, in what follows, we will show that
\begin{equation}%
\begin{array}
[c]{lll}%
\hat{P}(K,T)^2\ge \hat{P}(K-1,T)\hat{P}(K+1,T).
\end{array}
\label{eq:P(K,T) log-concave}
\end{equation}
Based on (\ref{eq:P(K,T) difference}), (\ref{eq:P(K,T) log-concave}) can be rewritten as
\begin{equation}%
\begin{array}
[c]{lll}%
\hat{P}(K-1,T)(\frac{T-K+1}{K})\frac{1-\epsilon}{\epsilon} - \hat{P}(K,T)\ge 0.
\end{array}
\label{eq:P(K,T) log-concave2}
\end{equation}
We use induction to show (\ref{eq:P(K,T) log-concave2}). For $T=1,2$, it is obvious to see that (\ref{eq:P(K,T) log-concave2}) holds. For $T=3$, we can verify (\ref{eq:P(K,T) log-concave2}) by using (\ref{eq:P(K,T) N=1}). Assume that for $T=t>3$, (\ref{eq:P(K,T) log-concave2}) holds. For $T=t+1$, after some algebra, we have
\begin{equation}%
\begin{array}
[c]{lll}%
&&\hat{P}(K-1,t+1)(\frac{t-K+2}{K})\frac{1-\epsilon}{\epsilon} - \hat{P}(K,t+1)\\
&=&\hat{P}(K-1,t)(\frac{t-K+1}{K})\frac{1-\epsilon}{\epsilon} - \hat{P}(K,t)\\
&+&\frac{1}{K}(\hat{P}(K-1,t)\frac{1-\epsilon}{\epsilon}-\binom{t}{K-1}\epsilon^{t-(K-1)}(1-\epsilon)^{K})
\ge 0,
\end{array}
\end{equation}
which is based on the induction and (\ref{eq:P(K,T) difference}).

2) Since $P(K_t,t)$ is monotonically increasing with $t$ for any $K_t$, $R_t(K_t)$ is monotonically increasing with $t$. Besides, as $t$ goes to infinite, $\lim\limits_{t\rightarrow\infty}R_t(K_t)=K_t$, i.e., the block with length $K_t$ can be delivered almost surely. Therefore, for any $K<\hat{K}_t$, we can conclude that $R_{t+1}(\hat{K}_t)> R_{t+1}(K)$. Since $\hat{K}_{t+1}$ is the optimal block size in slot $t+1$, i.e., $R_{t+1}(\hat{K}_{t+1})\ge R_{t+1}(\hat{K}_{t})$, if $\hat{K}_{t+1}<\hat{K}_{t}$, then we have $R_{t+1}(\hat{K}_{t+1})< R_{t+1}(\hat{K}_{t})$, which contradicts the fact that $\hat{K}_{t+1}$ is the optimal block size in slot $t+1$. Therefore, $\hat{K}_{t+1}\ge\hat{K}_{t}$.

%

\subsection{Proof of Theorem \ref{thm:decreasing}}\label{pf:thm:decreasing}
The proof follows from a contradiction argument. Suppose that $K_{t}^* > K_{t+1}^*$. It can be shown that $K_t^* \le \hat{K}_t$ by a contradiction argument. From Proposition \ref{prop:unimodal}, it follows that $R_t(K_{t}^*)>R_t(K_{t+1}^*)$ in slot $t$. Since $K_{t}^*$ is the optimal action in slot $t$, $V_t(K_{t}^*;\mathcal{P}^*)>V_t(K_{t+1}^*;\mathcal{P}^*)$. Since $K_{t}^* > K_{t+1}^*$
in slot $t$, when the optimal policy is applied, the future reward $J_t(K_{t}^*)$ under $K_{t}^*$ is less than the future reward $J_t(K_{t+1}^*)$ under $K_{t+1}^*$, due to the less remaining time under $K_{t}^*$. The future rewards under both actions are monotonically increasing with $t$, since $R_t(K)$ is monotonically increasing with $t$ for given $K$. Moreover, $\lim\limits_{t\rightarrow\infty}(J_t(K_{t}^*)-J_t(K_{t+1}^*))=0$, since the  probability of successfully delivering any given set of packets under any policy goes to 1, when the remaining time goes to infinity. This indicates that the gap between these future rewards decreases in slot $t+1$. Since $R_t(K)$ is monotonically increasing with $t$,
we have $R_{t+1}(K_{t}^*)>R_{t+1}(K_{t+1}^*)$. Therefore, $R_{t+1}(K_{t}^*)+J_{t+1}(K_{t}^*)>R_{t+1}(K_{t+1}^*)+J_{t+1}(K_{t+1}^*)$, i.e.,
in slot $t+1$, the total expected reward under $K_{t+1}^*$ is less than that under $K_{t}^*$, which contradicts that $K_{t+1}^*$ is the optimal action in slot $t+1$.

\subsection{Proof of Theorem \ref{thm:optimal<greedy}}\label{pf:thm:optimal<greedy}
The proof follows from a contradiction argument. Suppose that $K_t^* > \hat{K}_t$. For any sample path, the case with $\hat{K}_t$ will deliver the block earlier than the case with $K_t^*$. For the sample paths with the number of slots that all the channels between the transmitter and the receivers are good less than $K_t^*$, the reward under $\hat{K}_t$ is higher than that under $K_t^*$. For the other sample paths with the number of slots that all the channels between the transmitter and the receivers are good greater than $K_t^*$, the block with size $\hat{K}_t$ will be delivered earlier than that with size $K_t^*$. We assume that after the block with size $\hat{K}_t$ is delivered, the scheduler chooses to deliver the block with size 1, before the block with size $K_t^*$ is delivered. Then after the block with size $K_t^*$ is delivered, the optimal policy is applied for both cases. Obviously, in this case, both cases will generate the same reward. However, for the case with block size $\hat{K}_t$, the policy that we applied after the block with size $\hat{K}_t$ is delivered may not be optimal, which means that the reward under the optimal policy is no less than the reward of the policy we used. Therefore, the total expected reward under $K_t^*$ is less than that under $\hat{K}_t$, which contradicts the fact that $K_t^*$ is the optimal action in slot $t$.

\subsection{Proof of Corollary \ref{cor:Kt}}\label{pf:cor:Kt}
From Theorem \ref{thm:optimal<greedy}, we have $K_j^*\le\hat{K}_j$. Therefore, it suffices to show that $\hat{K}_j\le K_t$ for any $j\in\{1,...,t\}$. From Proposition \ref{prop:unimodal}, $K_t$ is in the decreasing sequence of $R_t(\cdot)$ when $R_t(K_t)>R_t(K_t+1)$. Therefore, it follows that $\hat{K}_t\le K_t$.

\subsection{Proof of Theorem \ref{thm:greedyoptimal}}\label{pf:thm:greedyoptimal}
When $R_t(1) > R_t(2)$ holds, $\hat{K}(t) = 1$, due to the unimodal property of $R_t(\cdot)$. Then, $K^*(t) = 1$ from Theorem \ref{thm:optimal<greedy}. Since $K^*(t)$ is non-decreasign with $t$ (Theorem \ref{thm:decreasing}), $K^*(t') = 1$ in the remaining slots $t' > t$, i.e., the plain retransmission policy is optimal. To show there exits a threshold $\epsilon^*$, we expand $R_t(1) > R_t(2)$ according to (\ref{eq:R_t}), where $R_t(1) = (1-\epsilon^T)^N$ and $R_t(2) = 2(1-\epsilon^T+T\epsilon^T-T\epsilon^{T-1})^N$. Then, the monotonicity of $\epsilon^*$ follows from comparing $R_t(1)$ with $R_t(2)$ in the expanded form. Define $f(\epsilon,t,N) = (1-\epsilon^t) - 2^{1/N}(1-\epsilon^t+t\epsilon^t-t\epsilon^{t-1})$ such that $f(\epsilon^*(t,N),t,N)=0$. Note that $f(0,t,N) = 1-2^{1/N} <0$ and $f(1,t,N) = 0$. There exists a unique non-trivial value of $\epsilon'$ in $(0,1)$ to maximize $f(\epsilon,t,N)$. For $\epsilon < \epsilon'$, $f(\epsilon,t,N)$ is first increasing and then decreasing back to $0$. Therefore, there exists a unique non-trivial solution of $f(\epsilon^*(t,N),t,N) = 0$ such that $f(\epsilon,t,N) < 0$ for $\epsilon < \epsilon^*(t,N)$ and $f(\epsilon,t,N) > 0$ for $\epsilon > \epsilon^*(t,N)$.

\subsection{Proof of Corollary \ref{cor:threshold}}\label{pf:cor:threshold}
If $N_2 > N_1$, $f(\epsilon,t,N_2) > f(\epsilon,t,N_1)$. For any $N$, $f(\epsilon,t,N)$ increases with $\epsilon$, achieves a positive maximum and decreases back to zero. Since $f(\epsilon,t,N_2) > f(\epsilon,t,N_1)$, the value $\epsilon^*_i(t,N_i)$, $i=1,2$, for which $f(\epsilon^*_i,t,N_i) = 0$ decreases from $\epsilon^*_1(t,N_1)$ to $\epsilon^*_2(t,N_2)$. By following the similar arguments, it follows that $\epsilon^*(t,N)$ is  monotonically increasing with $t$.


\end{document}